\documentclass{jfm}

\usepackage{graphicx}
\usepackage{amssymb}
\usepackage{amsmath}
\usepackage{hyperref}
\usepackage{cleveref}
\usepackage{bbold}
\usepackage{epstopdf}
\usepackage{verbatim}

\crefname{equation}{equation}{equations}
\crefname{figure}{figure}{figures}

\newcommand{\bms}[1]{\boldsymbol{#1}}
\newcommand{\nn}{\nonumber}
\newcommand{\vphi}{\varphi}
\newcommand{\veps}{\varepsilon}
\newcommand{\bn}{\boldsymbol{\nu}}

\newcommand{\rp}{R}

\newcommand{\fg}{g}
\newcommand{\fd}{g_D}
\newcommand{\fs}{g_S}

\newcommand{\mob}{\mu_e}
\newcommand{\ld}{\lambda_D}
\newcommand{\epsa}{\bar{\veps}}

\usepackage{lmodern,pdftexcmds}

\DeclareMathAlphabet{\mathsfbi}{OT1}{\sfdefault}{bx}{sl}
\DeclareMathVersion{sfletters}
\SetSymbolFont{letters}{sfletters}{OML}{ntxsfmi}{b}{it}

\makeatletter
\newcommand{\mathbfsbilow}[1]{%
  \text{\mathversion{sfletters}$\m@th#1$}%
}

\DeclareRobustCommand{\tensor}[1]{%
  \begingroup
  \ifcat\noexpand #1\relax
    \edef\greek@test{\detokenize{#1}}%
    \edef\greek@test{\expandafter\@cdr\greek@test\@nil}%
    \edef\greek@test{\expandafter\@car\greek@test\@nil}%
    \edef\x{\the\lccode\expandafter`\greek@test}%
    \edef\y{\number\expandafter`\greek@test}%
    \ifnum\x=\y\relax
      \mathbfsbilow{#1}%
    \else
      \mathsfbi{#1}%
    \fi
  \else
    \mathsfbi{#1}%
  \fi
  \endgroup
}
\makeatother

\begin{document}

\author{Sela Samin\aff{1}
\corresp{\email{S.Samin@uu.nl}}
\and Ren\'{e} van Roij\aff{1}}

\shorttitle{Solvo-osmotic flow}
\shortauthor{S. Samin, and R. van Roij}

\title{Solvo-osmotic flow in electrolytic mixtures}

\affiliation{\aff{1}Institute for Theoretical Physics, 
Center for Extreme Matter and Emergent Phenomena, 
Utrecht University, Princetonplein 5, 3584 CC Utrecht, The Netherlands}




\maketitle

\begin{abstract}
We show that an electric field parallel to an electrically \emph{neutral} 
surface can generate flow of electrolytic mixtures in small channels. We term 
this solvo-osmotic flow, since the flow is induced by the asymmetric 
preferential solvation of ions at the liquid-solid interface. The generated 
flow 
is comparable in magnitude to the ubiquitous electro-osmotic flow at charged 
surfaces, but for a fixed surface charge density, it differs qualitatively in 
its dependence on ionic strength. Solvo-osmotic flow can also be sensitively 
controlled with temperature. We derive a modified Helmholtz-Smoluchowski equation that 
accounts for these effects.
\end{abstract}

\section{Introduction}

The use of electric fields to drive flow parallel to charged surfaces at the 
micro- and nano-scale is 
ubiquitous in science and technology, with applications in numerous fields, from 
``Lab-on-a-Chip'' devices \citep{stone2004,squires2005} and nanofluidics 
\citep{bocquet2010} to membrane and soil science, and the separation and analysis 
of biological macromolecules \citep{eijkel2005}. In its 
simplest form, the fluid velocity $v_\infty$ far from the charged surface is 
proportional to the applied electric field $E_0$, $v_\infty=\mob E_0$. The electro-osmotic mobility $\mob$ is given by the 
Helmholtz-Smoluchowski equation \citep{smoluchowski1903}, $\mob=-\veps \zeta/\eta$, 
where $\veps$ is the solvent permitivity, $\eta$ is the solvent viscosity, and 
 the so-called zeta potential $\zeta$ is the electric potential at the 
shear plane. In neat (single-component) solvents, this 
electro-osmotic flow (EOF) has been studied 
extensively \citep{delgado2007}, with more recent works focusing on the 
influence of the channel geometry \citep{bhattacharyya2005,mao2014}, 
surface slip \citep{huang2008,bouzigues2008,maduar2015,rankin2016}, ion-specificity \citep{huang2007}
and the solvent rheology \citep{bautista2013}. 
EOF in solvent mixtures is common in non-aqueous capillary electrophoresis \citep{kenndler2014}, 
and was investigated also for aqueous mixtures \citep{valko1999,grob2002}, 
but the electrokinetics in such systems is not well understood.

The differences between the electrostatics of neat solvents compared to solvent 
mixtures has also been an area of intense research in recent years, with the 
growing use of miscible and non-miscible oil-water mixtures in colloidal 
science and microfluidics.
A crucial role is played by the partitioning of ions between solvents, due to their preferential solvation in one of the liquids 
\citep{onuki:3143,tsori_pnas_2007,zwanikken2007,ben-yaakov2009,
araki2009,efips_epl,
samin2013,onuki2011,bier2011,pousaneh2011,efips_jcp_2012,bier2012,pousaneh2014,
michler2015,samin2016a} .
The preferential wetting of one liquid at a solid surface or the presence 
of a liquid-liquid interface therefore also affects the electrostatics of the 
mixture. It 
leads to a modification of colloid-colloid 
\citep{Beysens1998,Bonn2009,
Bechinger2008,
nellen2011,samin2014} as well as colloid-interface interactions 
\citep{leunissen2007,leunissen2007b,elbers2016,banerjee2016,everts2016}. The 
strength of preferential solvation is measured by the Gibbs 
transfer energy $k_BT\fg^\alpha$ of an ion species $\alpha$ between two 
solvents, where $k_BT$ 
is the thermal energy, and $|\fg^\alpha|\sim 1-10$ for aqueous mixtures of relatively polar organic solvents \citep{marcus_cation,marcus_anion}, 
but can be as large as $15$ in less polar solvents containing antagonistic salts \citep{onuki2016}. 
For monovalent salts dissolved in mixtures, the combined effect of both ionic 
species 
\citep{samin2013,onuki2011,bier2011,michler2015} is conveniently expressed in 
terms of the average overall solubility $\fs=(\fg^++\fg ^-)/2$, and the 
solubility contrast, $\fd=(\fg^+-\fg^-)/2$, which for example, determines the 
Donnan potential $k_BT\fd/e$ at oil-water interfaces due to ion partitioning, 
where $e$ is the elementary charge. In this work, we show 
that preferential solvation also affects electrokinetic phenomena in mixtures, 
where it generates an additional, and significant, source of fluid mobility. 
We find that the dominant contribution to this additional mobility is 
proportional to $\fd$ and is independent of the surface charge, and therefore 
able to generate flow even at electrically neutral surfaces.

\section{Formulation of the problem}

Within a continuum theory, we study the flow in a channel containing an 
electrolytic mixture using direct 
numerical simulations and a simple linear theory. We consider an oil-water 
mixture characterized by the order parameter $\vphi$, which is the deviation of 
the volume fraction of water in the mixture, $\phi$, from its critical value $\phi_c$: $\vphi=\phi-\phi_c$. The mixture 
contains point-like monovalent ions with number densities 
$n^\pm$. In our calculations, two cylindrical reservoirs containing the 
mixture are coaxially connected by a long cylindrical channel with radius $\rp$, with the $z$ axis being the axis of symmetry and $r$ the radial coordinate. In 
equilibrium, the composition 
in both the reservoirs is $\vphi_0=0$ and the number densities of ions are $n_0$. At the edge of one reservoir, we impose a uniform external 
electric field $E_0\hat{z}$, forcing the charged mixture in the channel, 
and leading to a flow field $\bms{v}$. 

\subsection{Governing equations}

To study the mixture dynamics we start from a free energy $F$ of the form 
\begin{align}
\beta F&=\int {\rm d} \bms{{\rm r}} ~ \Bigl\{f_m(\vphi)+\frac{C}{2}|\bnabla 
\vphi|^2 +\frac{\beta}{2}\veps(\vphi) (\bnabla
\psi)^2 \nn \\ &+\sum_{\alpha=\pm} n^\alpha\left[\left(\log
(v_0n^\alpha)-1\right)-\fg^\alpha\vphi \right] \Bigr\},
\label{eq:fe}
\end{align}
where $\beta^{-1}=k_BT$ and $v_0=a^3$ is the molecular 
volume of both mixture components. The first term in the integrand is the 
``double-well'' bulk mixture free energy 
density, $v_0f_m=(\chi-2)\vphi^2+4\vphi^4/3$, where $\chi\sim1/T$ 
is the Flory parameter. 
This free energy leads to an 
upper critical solution temperature type phase diagram, with a 
critical temperature $T_c$ and the corresponding critical Flory parameter 
$\chi_c=2$. In this work, we focus on the region $T>T_c$ ($\chi<\chi_c$), such 
that the bulk mixture is always homogeneous.
The second ``square gradient'' term accounts for 
composition 
inhomogeneities at interfaces, where $C=\chi/a$. The third term 
is the electrostatic energy density, where $\psi$ is the electric potential, and $\veps(\vphi)$ is the permitivity, assumed to depend linearly on 
$\vphi$. The final term in the integrand is the ionic free energy, 
composed of 
the ideal-gas entropy of ions of species $\alpha=\pm$, and the ionic solvation 
energy, which is 
proportional to the Gibbs free energy of transfer $\fg^\alpha$ and local 
solvent composition. Note that $\fg^\alpha$ can greatly vary 
between ionic species depending on their size, charge and chemistry.  

The relations governing the mixture dynamics read \citep{araki2009}
\begin{align}
\label{eq:ch}
 \partial \vphi/\partial t + \bnabla \cdot \left( \vphi \bms{v} \right) 
&=
D^m \nabla^2 \beta\mu~,\\
\label{eq:pn}
  \partial n^\alpha /\partial t + \bnabla \cdot \left( n^\alpha \bms{v} \right)
&= D^\alpha \bnabla \cdot n^\alpha \bnabla \beta\lambda^\alpha, \\
\label{eq:pos}
\bnabla \cdot (\veps(\vphi) \bnabla \psi ) &=-\textstyle\sum_\alpha z^\alpha 
n^\alpha,\\
\label{eq:nc}
\bnabla \cdot \bms{v}&=0, \\
\eta \nabla^2 \bms{v}-\bnabla p&= v_0^{-1}\vphi \bnabla \mu + 
\textstyle\sum_\alpha n^\alpha \bnabla 
\lambda^\alpha ~.
\label{eq:ns}
\end{align}
\cref{eq:ch} is the convective Cahn-Hilliard equation, where $D^m$ is the 
inter-diffusion constant of the mixture and $\mu=v_0(\delta F /
\delta \vphi $) is the solvent
chemical potential given by $\beta\mu/v_0= -C
\nabla^2\vphi+f'(\vphi) 
-\beta\veps'(\vphi)(\bnabla\psi)^2/2-\sum_\alpha 
\fg^\alpha n^\alpha$. \cref{eq:pn,eq:pos} are the Poisson-Nernst-Planck 
equations, where $D^\alpha$ are ionic diffusion 
constants and $\lambda^\alpha=\delta F /
\delta n^\alpha $ are the ionic
chemical potentials given by $\beta\lambda^\alpha= 
\log(v_0n^\alpha)+\beta z^\alpha  
\psi-\fg^\alpha\vphi$, with $z^\alpha=\pm e$. \cref{eq:nc,eq:ns} are the 
Stokes equations at small 
Reynolds number, where $p$ is the pressure and the right hand side of 
\cref{eq:ns} contains the body forces due to concentration gradients. 

\begin{figure}
\centering
\includegraphics[width=3.5in,clip]{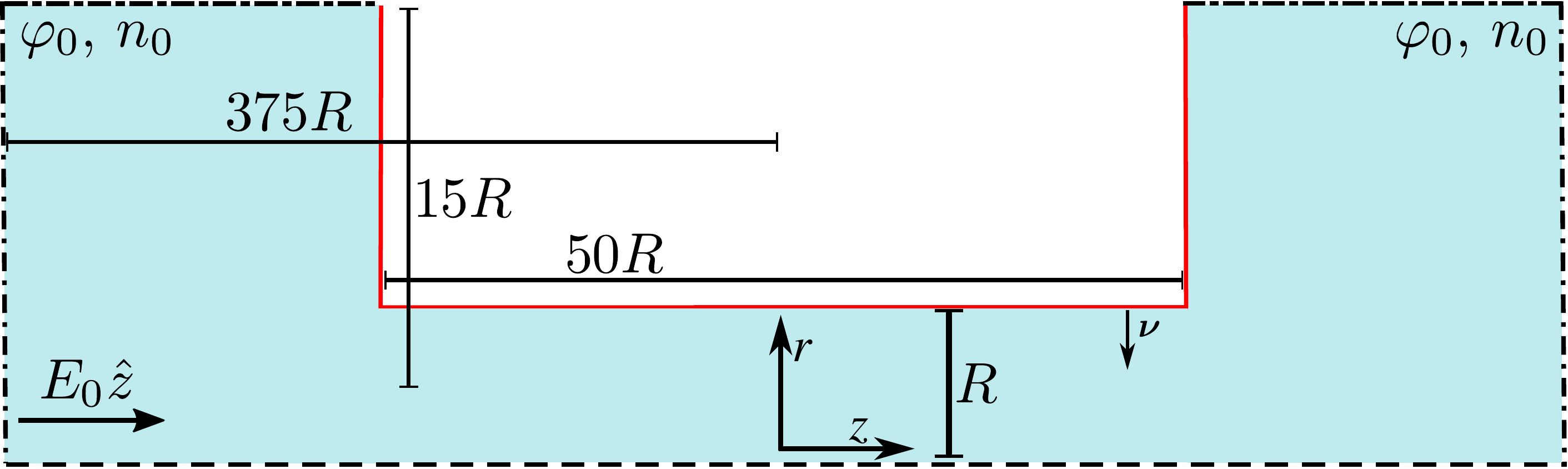}
\caption{An illustration of the system in the $rOz$ plane. A cylindrical channel with 
radius $R$ connects two reservoirs containing a salty water--acetonitrile mixture with a composition $\vphi_0=0$ and a salt concentration $n_0$. The system is driven out of equilibrium by 
an external electric field $E_0$ in the $\hat{z}$ direction. The system dimensions, given in the figure in terms of $R$, ensure that the effect of reservoirs and channel edges on the flow that develops in the channel is small. Different boundary line types indicate different boundary conditions (see text).}
\label{fig_sys}
\end{figure}
%


An illustration of the computational domain for the numerical solution of 
\cref{eq:ch,eq:pn,eq:pos,eq:nc,eq:ns} is shown in 
\cref{fig_sys} (not to scale). Solid walls are indicated by 
the red lines in \cref{fig_sys}. On these walls, we impose the no-slip boundary 
condition (BC) for the velocity, $\bms{v}=0$, and no material fluxes for the 
composition and ions. The second BC for the composition is $-\bn\cdot\bnabla 
\vphi=\gamma$, with $\bn$ being the unit normal to the surface. The 
coefficient $\gamma$ represents the short-range interaction between the solvent 
mixture and the surface (per solvent molecule). We call $\gamma$ the 
effective surface field; $\gamma$ is positive (negative) for hydrophilic 
(hydrophobic) surfaces. The surface may carry a fixed charge density $e\sigma$ which 
from Gauss's law implies $-\bn\cdot\bnabla \psi=e\sigma/\veps(\vphi)$.

At the open reservoir edges, indicated by dash-dot lines in 
\cref{fig_sys}, we set the composition to $\vphi_0=0$ and ion densities to 
$n_0$. We also allow the mixture to be freely advected, with vanishing total 
stress and diffusive fluxes:  $-\bn\bcdot\left(p\tensor{1}+\tensor{S}-\tensor{T} 
\right)=0~$, $ -\bn \bcdot \bnabla \mu=0$ and $-\bn \bcdot n^\pm\bnabla 
\lambda^\pm=0$, respectively, where 
$\tensor{T}=\eta(\bnabla\bms{v}+\bnabla\bms{v}^T)$ is the viscous stress 
tensor and $\tensor{S}$ is the 
total stress tensor given by
\begin{align} 
\beta\tensor{S} &=\Bigl[ 
\vphi f_m'(\varphi) -f_m-\tfrac{C}{2}
|\bnabla\vphi|^2-C\vphi\nabla^2\vphi-\vphi\veps'(\vphi)\frac{\left|\bms{E}
\right|^2 } { 2 } +\frac {
\bms{ E}\cdot\bms{D}}{2} \nonumber \\ &+\sum_ { \alpha=\pm } (1-\fg^\alpha\vphi)n^\alpha 
\Bigr] \tensor{1}+C\bnabla\vphi\bnabla\vphi-\bms{E}\bms{D}~,
\end{align} 
where $\bms{E}=-\bnabla \psi$ is the electric field, $\bms{D}=\veps(\vphi) \bms{E}$ is the displacement field, and the prime 
indicts differentiation with 
respect to the argument. Lastly, the dashed curve in \cref{fig_sys} is the $r=0$ axis where we apply symmetry BCs for all fields. Numerical simulations in this work were performed using the finite-elements software COMSOL 
multiphysics.

\subsection{Linear theory}

To better understand the flow generated in the channel, we first consider a 
simplified system where we assume that: (i) perturbations in the composition 
$\vphi-\vphi_0$ and ion densities $\delta n^\pm = n^\pm-n_0$ relative to their 
bulk values are small due to weak adsorption at the channel surface, (ii) the 
channel is very long such that edge effects are negligible and translational invariance in the $z$ direction applies, (iii) the dominant 
body force in \cref{eq:ns} is the electric body force, and (iv) convective 
composition and ion currents in the channel are negligible since the 
corresponding P\'{e}clet numbers are small. However, even in simple mixtures, 
transport through channels can lead to complex and surprising 
effects \citep{samin2017}, also for small P\'{e}clet numbers. 
Therefore, it is necessary to also solve the complete 
system \cref{eq:ch,eq:pn,eq:pos,eq:nc,eq:ns} numerically to verify that the simplified description is valid. 

Within the appropriate parameter regime, the full transport problem can thus be 
reduced to a steady-state problem, where all fields depend only on $r$. We also 
decompose the electric potential as $\psi(r,z)= \Psi(r)/(\beta e)-zE_z+const$, 
where $\Psi(r)$ is a dimensionless radial potential and $E_z$ is the 
\emph{uniform} axial electric field in the channel. Note that $E_z \neq E_0$ in 
general. For a 
fully developed flow, we assume $v_r=0$ and thus 
$v_z=v_z(r)$ from \cref{eq:nc}. Lastly, we set 
$\bnabla p=0$ as we focus on the effect of $E_z$. Linearizing 
the simplified \cref{eq:ch,eq:pn,eq:pos,eq:nc,eq:ns} \citep{onuki2011,samin2013} 
we find $\delta n^\pm=n_0(\fg^\pm\vphi\mp\Psi)$ and obtain the system:
\begin{align}
\label{eq:lin_ch}
\xi^2\nabla_r^2\vphi&=\vphi+ 2n_0\fd(\Psi- \fd\vphi)/\tau~,\\
\label{eq:lin_pos}
\ld^2\nabla_r^2\Psi&=\Psi-\fd \vphi~,\\ 
\label{eq:lin_ns}
\ld^2\eta \nabla_r^2 v_z&=\epsa (\Psi-\fd \vphi)E_z/(\beta e)~,
\end{align}
where $\xi=(C/\tau)^{1/2}$ is the modified correlation length and $\tau 
=2(2-\chi)/v_0-2n_0\fs^2$. The Debye length is $\ld=(8\pi l_Bn_0)^{-1/2}$, 
where 
$l_B=e^2/(4\pi\epsa k_BT)$ is the Bjerrum length with the average permitivity 
$\epsa=\veps(\vphi_0)$. 

\section{Results}

\begin{figure}
\centering
\includegraphics[width=3.3in,clip]{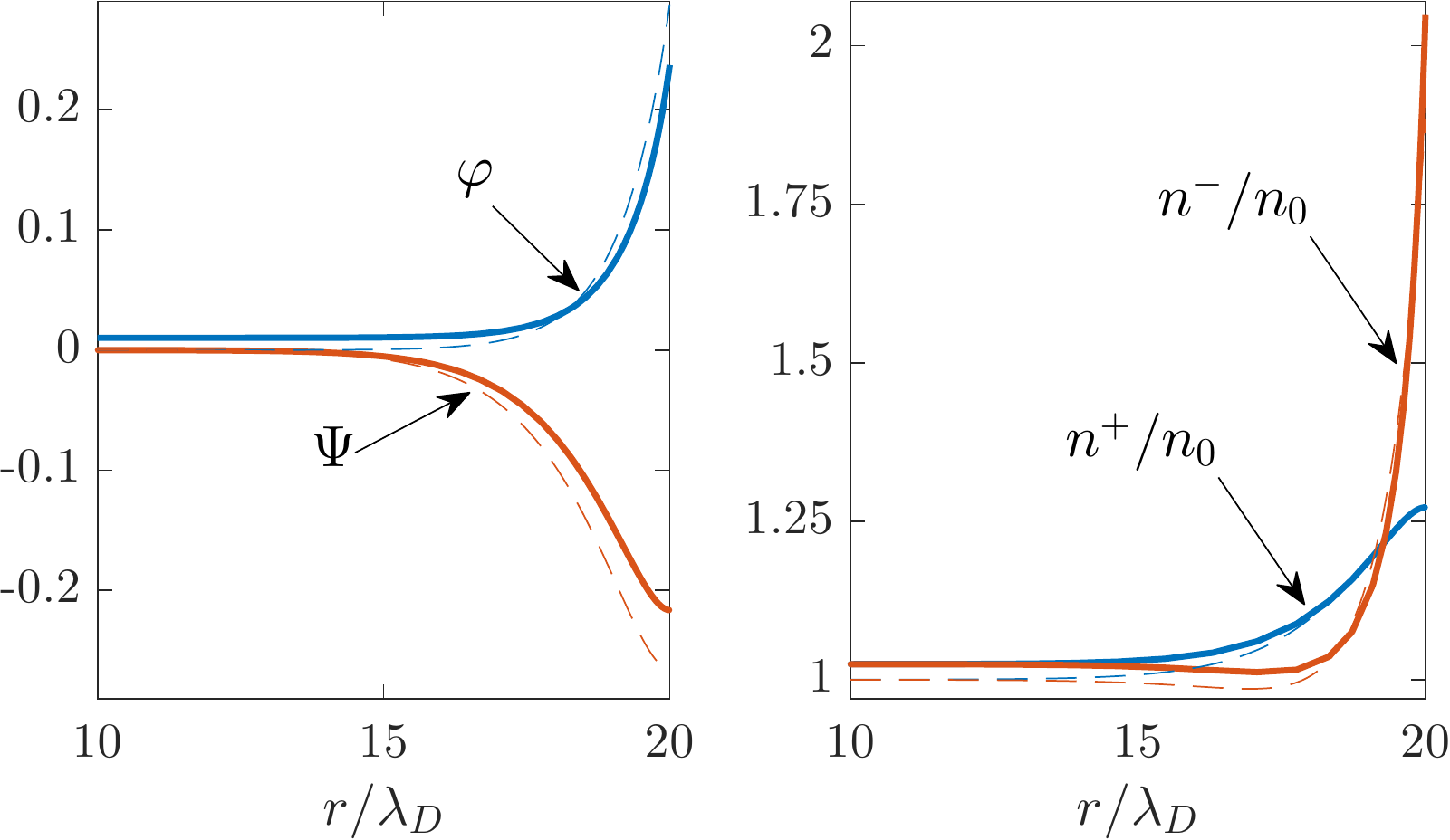}
\caption{(a) Composition $\vphi(r)$ and scaled potential $\Psi(r)$ at the 
channel center ($z=0$) and (b) the corresponding scaled cation and anion 
profiles, for a water--acetonitrile \citep{sazonov2007,wohlfarth2009} with 
a Debye 
length $\ld=1.16$ nm and a correlation length $\xi=1.31$ nm. Solid lines: 
numerical solution, dashed curves: linear 
theory (see text). We plot the profiles near the channel surface, where here 
and in all other figures the channel radius is $\rp=20\ld$. The channel surface 
is uncharged ($\sigma=0$) and 
hydrophilic, $\gamma=0.2C$, corresponding to a tension of $\approx7$ mN/m. The 
mixture with $\epsa=58$ and at a temperature of $293$ K ($\chi=1.86$) contains $50$ mM of 
NaCl ($\fg^+=0,\fg^-=4$). For the mixture properties we use the critical 
temperature $T_c=272$ K, a molecular length $a=3.6$ \AA{}, estimated from a 
critical density of 900 kg/m$^3$, and a viscosity of $0.7$ mPa s 
\citep{sazonov2007,wohlfarth2009}. The 
liquid permeability is given by $\veps(\vphi)/\veps_0=\veps_{\rm 
acetonitrile}/2+\veps_{\rm water}/2+\left(\veps_{\rm 
water}-\veps_{\rm acetonitrile}\right)\vphi$, where $\veps_0$ is the 
vacuum permitivity, and we used $\veps_{\rm acetonitrile}=36.6$ and $\veps_{\rm 
water}=79.5$.  
}
\label{fig_prof1} 
\end{figure}

In a cylindrical domain, the profiles $\vphi(r)$ and $\Psi (r)$ of 
\cref{eq:lin_ch,eq:lin_pos} are the linear 
combinations
\begin{align}
\label{eq:solpb}
 \Psi (r)&=a_1 \mathrm{I}_0(q_1 r) -a_2 \mathrm{I}_0(q_2 r)~,\\
\label{eq:solcomp}
  \vphi(r)&=b_1 \mathrm{I}_0(q_1 r) -b_2 \mathrm{I}_0(q_2 r)~,
\end{align}
where $\mathrm{I}_n$ is the modified Bessel function of the first kind of order $n$, 
such that the wave numbers $q_i$ ($i=1,2$) obey the biquadratic equation
\begin{equation}
 \label{eq:solcond1}
  \ld^2 q_i^4-\left[1+(\ld/\xi)^2-\fd^2/(4\pi l_B 
C)\right]
q_i^2+\xi^{-2}=0~.
\end{equation}
The amplitudes $a_i$ and $b_i$ follow from symmetry at $r=0$ and the BCs 
given above at the channel surface $r=\rp$.  Making also use of the identity 
$((q_1\ld)^2-1)((q_2\ld)^2-1)=\fd^2/(4\pi l_BC)$, which follows from 
\cref{eq:solcond1}, the amplitudes read:
\begin{align}
\label{eq:ai}
 a_i=&  \frac{ \fd}{
1-(\ld q_i)^2}b_i~,\\
\label{eq:bi}
 b_i=&\frac{\gamma\left[1-(\ld q_i)^2\right]-  \fd\sigma/C}{
\ld^2 q_i(q_2^2-q_1^2)\mathrm{I}_1(q_i
\rp)}~.
\end{align}


We plot in \cref{fig_prof1}a the resulting composition and potential profiles for a 
water--acetonitrile mixture with $n_0=50$ mM of NaCl ($\ld=1.16$ nm, 
$\fg^+\approx 0$ and 
$\fg^-\approx 4$ \citep{marcus_cation, marcus_anion}) at room 
temperature ($\chi=1.86$, $\xi=1.31$ nm) inside a channel with radius 
$R=20\ld$. The mixture physical properties are given in the caption of 
\cref{fig_prof1}. We assume that these properties are independent of 
temperature, except for the mixture inter-diffusion constant $D^m$, which 
follows from $D^m=(6\beta\pi\eta\xi)^{-1}$ \citep{kawasaki1970}. The ionic 
diffusion constants are taken to be $D^\pm=1\times 10^{-9}$ m$^2$/s.

The channel surface is uncharged ($\sigma=0$), but is hydrophilic 
($\gamma=0.2C>0$), resulting 
in the adsorption of water near the wall. Since the anions 
are hydrophilic 
the water ``drags'' them along, and hence the anions also effectively adsorb at 
the surface, see the ionic profiles in \cref{fig_prof1}b. Although the 
cations are indifferent to the local composition, they 
also adsorb at the surface due to electrostatics, but are distributed more 
broadly. 
The result is an electric double layer at an uncharged surface 
\citep{samin2013,samin2014}, even though 
this layer is overall charge neutral. \cref{fig_prof1} shows a good 
agreement between the full numerical solutions and the linear theory. The 
numerical profiles of $\vphi$ and $n^\pm$ decay to values 
that slightly differ from their bulk values due to a weak nonlinear effect 
induced by solvation \citep{samin2016a}. The thickness of the charged layer at the surface $\rho(r)=2e n_0(\fd\vphi-\Psi)$ could 
be estimated in experiments from the characteristic lengths $q_i^{-1}$. These length scales are to be estimated from \cref{eq:solcond1} given data for $\ld$, $\xi$ and $g_D$.

\begin{figure}
\centering
\includegraphics[width=3.3in,clip]{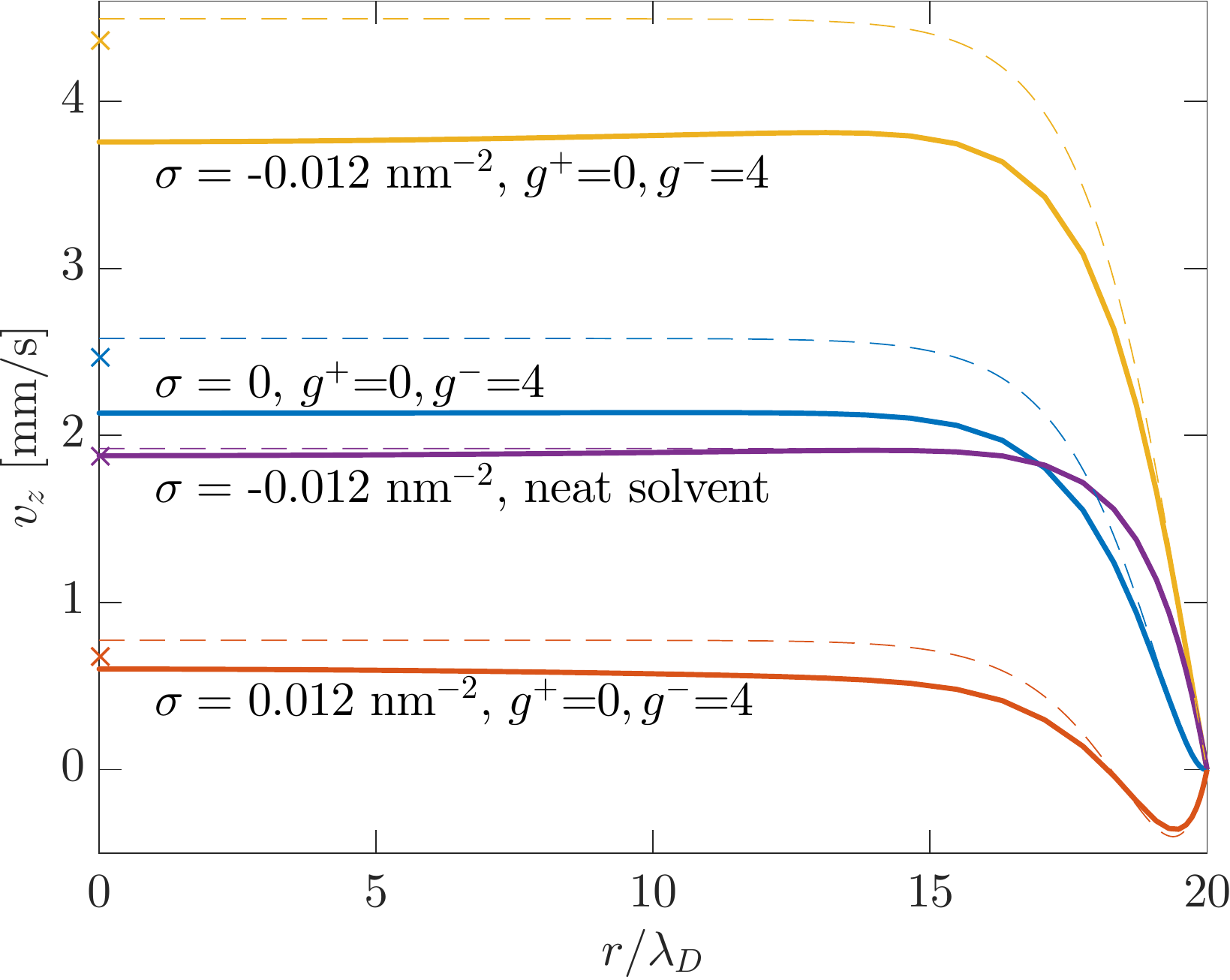}
\caption{Axial velocity profiles $v_z(r)$ at an applied field of $E_0=2.5$ 
V/mm for several scenarios, see the labels in the figure. All other parameters 
are the same as in \cref{fig_prof1}. Numerical results (solid curves) are 
in good agreement with linear theory (dashed curves), where in 
\cref{eq:ns_lin_sol} we used $E_z$ from the numerical calculation. Crosses 
were calculated using \cref{eq:lin_mu}. }
\label{fig_prof2} 
\end{figure}

Although the fluid is overall neutral, the negative ions layer is more strongly adsorbed at the wall. When an external axial electric field is applied, this layer then serves as an effective surface charge, and the adjacent positively charged layer as an effective electric double layer. The result is an induced fluid flow, similar to EOF. The steady-state axial flow profile $v_z(r)$ is obtained by putting \cref{eq:solpb,eq:solcomp} into 
\cref{eq:lin_ns}, and imposing the no-slip BC at $r=\rp$ and symmetry at 
$r=0$,
\begin{align}
\label{eq:ns_lin_sol}
v_z(r)=\frac{\epsa E_z}{\beta e \eta}\sum_{i=1,2} 
\left(\fd b_i-a_i\right)\frac{\mathrm{I}_0(q_i r)-\mathrm{I}_0(q_i \rp)}{(-1)^i (\ld q_i)^2}~,
\end{align}
which for the parameters of 
\cref{fig_prof1} is plotted in \cref{fig_prof2} (labeled with $\sigma=0$), 
showing reasonable agreement between the numerical solution and 
\cref{eq:ns_lin_sol}. Indeed, a flow is generated by the overall neutral but 
locally charged layer of fluid near the wall. \Cref{fig_prof2} shows that 
$v_z(r)$ quickly saturates to a plug-like flow of a few mm/s. We term this 
solvation-induced phenomenon \emph{solvo-osmotic flow} (SOF). This flow is 
similar and comparable in magnitude to simple EOF, 
shown in \cref{fig_prof2} by the curve for a neat solvent with the same 
properties as the mixture, but near a weakly charged surface with $\sigma=-0.012$ 
nm$^{-2}$.

SOF can 
either combine or work against EOF, as shown by two 
curves in \cref{fig_prof2}, where we used the parameter of \cref{fig_prof1}, 
but with non-zero surface charges, $\sigma=\pm0.012$ nm$^{-2}$. For 
$\sigma<0$ the result is an enhanced flow with the velocity at the channel 
center slightly smaller than the sum of SOF and EOF on their own. In a neat 
solvent, when $\sigma$ changes sign one expects a simple change of 
sign also for the velocity. However, here SOF opposes EOF for $\sigma>0$ and, 
for the 
parameters of \cref{fig_prof2}, the result is that inside most of the channel 
$v_z$ remains positive, but with a much smaller magnitude. There is, however, a 
weak back flow close to the channel wall resulting from the radial 
component of the body force in the fluid becoming significant in this case.

In the large-$R$ limit where $\rp\gg\ld,\xi$, the fluid moves as 
a 
plug, $v_z=const.$, and one defines the electro-osmotic mobility of the fluid 
as $\mob=v_z/E_z$. In this limit we find: 
\begin{align}
\label{eq:vr1}
\mob=-\frac{\epsa}{\eta}\frac{4\pi 
l_B\sigma 
(q_1^2+q_2^2+q_1q_2-\ld^{-2})+\fd\gamma \ld^{-2}}{q_1q_2(q_1+q_2)\beta e}~.
\end{align}
Noting that the biquadratic \cref{eq:solcond1} also implies 
$q_1q_2=(\xi\ld)^{-1}$ and $q_1^2+q_2^2=\xi^{-2}+\ld^{-2}\left[1-\fd^2/(4\pi 
l_BC)\right]$, and assuming that also 
$\ld/\xi\gg|\fd|/(4\pi l_BC)$, \cref{eq:vr1} reduces to the 
modified Helmholtz-Smoluchowski equation 
\begin{align}
\label{eq:lin_mu}
\mob=-\frac{\epsa}{\eta}\left[\left(4\pi l_B\ld 
\sigma+\frac{\fd\gamma\xi}{
1+\ld/\xi}\right)\frac{1}{\beta e} 
\right ]~.
\end{align}
\Cref{eq:lin_mu} is the main result of this paper, where the term in 
brackets is 
the zeta potential $\zeta$, with the shear plane located at $r=\rp$. Only the 
first term $\propto \sigma$ in \cref{eq:lin_mu} 
exists for a neat solvent \citep{keh2001}. In a mixture, $\zeta$ is determined also by 
the properties of the salt and surface through the second term 
$\propto\fd\gamma$ in 
\cref{eq:lin_mu}, which plays the role of an effective 
surface charge. The attractive feature of \cref{eq:lin_mu} is the ability to 
easily tune the fluid mobility through $\fd$ without making any surface 
modification. For example, by replacing the anion Cl$^{-}$ with the 
hydrophobic anion BPh$_4^{-}$, we have $\fd\approx6$ 
\citep{marcus_anion} (instead of $\fd=-2$), which would then change $\mob$ 
drastically, in some cases even its sign. As we 
show below, the dependence of $\mob$ on $\xi$ in mixtures opens the possibility 
to tune the mobility also with temperature, since $\xi$ diverges near the 
mixture critical temperature. Nevertheless, we stress that $\xi\sim a$ for 
completely miscible mixtures (the $\chi\rightarrow0$ limit in our theory). 
Hence, \cref{eq:lin_mu} should be applicable to mixtures in 
general, and as the crosses in 
\cref{fig_prof2} confirm, it is in good agreement with the numerical results 
for $v_z(0)$.

\begin{figure}
\centering
\includegraphics[width=5.25in,clip]{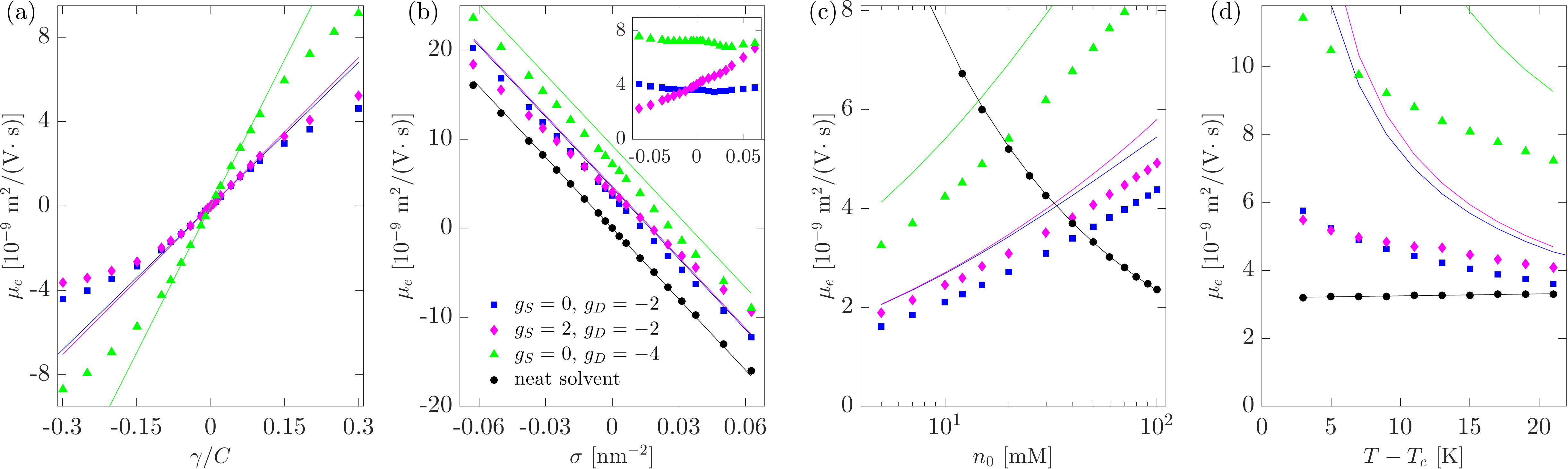}
\caption{Dependence of fluid mobility $\mob$ on (a) the surface-solvent 
coupling $\gamma$, (b) the 
surface charge $e\sigma$, (c) the salt concentration $n_0$ and (d) the 
temperature $T$, for three 
types of electrolytic mixtures (see text). Symbols correspond 
to 
numerical results and lines to \cref{eq:lin_mu}.
In (b)-(d), we plot also $\mob$ for a 
neat solvent 
with the 
same physical properties as the mixtures (circles). In panels (a)-(c), 
$T=293$ K is fixed while in panels 
(a), (b) and (d) we keep $n_0=50$ mM fixed. For the mixture data sets in 
(b)-(d) we 
took $\gamma=0.2C$ and in (a), (c) and (d) we took $\sigma=0$. For the 
neat solvent in (c)-(d), we used $\sigma=0.012$ nm$^{-2}$. Inset of (b): the 
shift in mobility between the mixtures and the neat solvent.}
\label{fig_prm1} 
\end{figure}

In \cref{fig_prm1} we show the effects of varying some of the free
parameters of our model on the fluid mobility, and compare numerical results 
(symbols) with \cref{eq:lin_mu} (lines). We compare 
three types of electrolytic mixtures, one containing NaCl (diamonds) as before 
and two 
containing so-called antagonistic salts in which one of the ions is hydrophilic 
and the other hydrophobic. In order to single out the role of $\fd$, we choose 
$\fs=0$ by setting $\fg^-=-\fg^+$=2 (squares) and 4 (triangles).
\cref{fig_prm1}a shows the pure SOF mobility as a function of the surface 
field $\gamma$, revealing that the linear theory becomes an 
excellent approximation for small enough $|\gamma|$, for which $\vphi\ll1$ 
holds, and that $\mob$ increases with $\gamma$ and $\fd$, as expected from 
\cref{eq:lin_mu}. Notice that the sign of $\mob$ depends on the wetting 
properties of the surface. For the salts having the same solubility 
contrast $\fd$ but different overall solubility $\fs$ (squares and diamonds), 
we find significant differences in the mobility only for large enough 
$|\gamma|$. In this regime, the ions' entropy decrease near the wall 
hinders their adsorption, leading to a reduced mobility compared 
to the linear theory.

In \cref{fig_prm1}b-d we also contrast the electrolytic mixtures with a 
neat 
solvent (circles) having the same physical properties as the mixtures. 
In \cref{fig_prm1}b we examine the effect of adding a fixed surface charge 
$\sigma$. Clearly, here SOF increases the mobility compared to regular 
EOF in a neat 
solvent.
The inset of \cref{fig_prm1}b shows that the shift in 
mobility is almost constant and increases with $\fd$ for the antagonistic 
salts, as expected from \cref{eq:lin_mu}. The mobility shift is not constant 
for NaCl, rather it increases with $\sigma$, and the result is an an asymmetry 
in the mobility when $\sigma$ changes sign, an effect that is absent in neat 
solvents. This effect originates from the nonlinear coupling between 
electrostatics and solvation \citep{efips_jcp_2012}, and is expressed by the 
next order term $\propto\fs\vphi\Psi$ in the expansion of $\delta 
n^\pm$. This contribution vanishes for our ideal antagonistic salts, but should 
be relevant for the majority of realistic salts.
The relative shift in 
mobility due to SOF is significant for the weakly to moderately charged 
surfaces we considered here, which is a common situation for hydrophobic 
surfaces \citep{jing2013}, or in hydrophilic surfaces near their iso-electric point. For the typically highly charged hydrophilic 
surfaces the contribution 
of SOF may be small, although this may not always be the case since both $\fd$ 
and $\gamma$ can be much larger than the values used in this work.

In \cref{fig_prm1}c-d we show the influence of the salt concentration 
and temperature, respectively, on mobility. In both panels we compare 
electrolytic 
mixtures in an uncharged channel with a neat solvent in a channel with  
$\sigma=0.012$ nm$^{-2}$. Strikingly, \cref{fig_prm1}c shows that the 
mobility increases with $n_0$ for the mixtures, 
while it decreases for 
the neat solvent. This behaviour follows from the dependence of $\mob$ on 
$\ld\propto n_0^{-1/2}$ in \cref{eq:lin_mu}. For constant 
$\sigma$, the contribution to the surface potential from the first term in 
\cref{eq:lin_mu} becomes smaller when $\ld$ decreases, while the second 
term becomes larger. \Cref{fig_prm1}c suggests that for $n_0\sim100$ mM, 
the contribution to the mobility from solvation could be dominant when the 
surface is weakly charged. SOF is enhanced by increasing the ionic strength since this leads also 
to increased ion adsorption at the wall, which results in a larger effective surface charge. 
The opposite dependence on $n_0$ between the two terms in \cref{eq:lin_mu} can therefore lead to a minimum in 
the mobility of mixtures at charged surfaces, which is confirmed by numerical calculations (not shown). 

\Cref{fig_prm1}d demonstrates that, in mixtures, temperature can be used to 
effectively tune the mobility, whereas its effect is much smaller in neat 
solvents. The reason is that the correlation length $\xi$ diverges as the 
$T_c$ is approached from the one-phase region at the critical composition. In
\cref{eq:lin_mu}, the mobility increases with $\xi$, or equivalently, when the 
temperature is lowered towards $T_c$, which is confirmed by the numerical 
results in
\cref{fig_prm1}d. Here, the agreement with the linear theory is not as good 
since $\vphi$ is no longer small when $T_c$ is approached. The numerical 
results show that $\mob$ changes by as much at $60\%$ in a temperature window 
of $20$ K, whereas it essentially remains constant for the neat solvent. 
We did not account, however, for the change of $\veps$ with 
temperature. This effect will be at play for both mixtures 
and neat solvents, and should be smaller since, for example, the permitivity 
of water changes only by about $10\%$ in the same temperature window.

\section{Conclusions}

In conclusion, we have shown that the electro-osmotic mobility of mixtures 
contains an additional contribution due to the preferential solvation of ions. 
This contribution allows to drive flow at 
an electrically neutral surface and to vary the mobility significantly by 
changing the type of salt or the temperature. All of our results can be directly 
extended to the phoretic motion of particles in an applied field.

A comparison of our results with existing experiments \citep{valko1999,grob2002} 
is difficult to preform at this point, because capillary electrophoresis experiments employ fused silica surfaces, 
which are typically highly-charged and, more importantly, charge-regulating. A modification of the theory is thus required to better mimic experimental conditions. This should include the chemical dissociation equilibrium at the channel surface, and the change in the fluid viscosity with composition. For the study of transient SOF, one should also consider the asymmetry in the cation and anion diffusion constants, and their dependence on the solvent composition. Our work points towards promising
possibilities for the utilization of SOF in the transport of fluids through 
channels and of colloids by external fields, and we hope it will lead to systematic experiments on the electrokinetics of solvent mixtures.

\begin{acknowledgments}
We acknowledge discussions with H. Burak Eral. We thank the anonymous referees for their useful comments and suggestions.
R.v.R acknowledges 
financial support of a Netherlands Organisation for Scientific Research (NWO) 
VICI grant funded by the Dutch Ministry of Education, Culture and Science 
(OCW). S.S acknowledges funding from the 
European Union's Horizon 2020 programme under the Marie 
Sk\l{}odowska-Curie grant agreement No. 656327. This work is part of the D-ITP 
consortium, a program of the NWO funded by the OCW.
\end{acknowledgments}

\bibliographystyle{jfm}

\end{document}